\documentclass[fleqn,usenatbib]{mnras}

\usepackage{newtxtext,newtxmath}
\usepackage[T1]{fontenc}

\DeclareRobustCommand{\VAN}[3]{#2}
\let\VANthebibliography\thebibliography
\def\thebibliography{\DeclareRobustCommand{\VAN}[3]{##3}\VANthebibliography}

\usepackage{graphicx}	
\usepackage{amsmath}	
\usepackage{ulem}
\usepackage{svg}

\newcommand{\hvec}[1]{\hat{\mathbf{#1}}}
\newcommand{\quotes}[1]{``#1''}

\title[Particle acceleration with MR in large scale RMHD simulations - I]{Particle acceleration with Magnetic Reconnection in large scale RMHD simulations: I. Current sheet identification and characterization}

\author[M. Nurisso et al.]{Matteo Nurisso$^{1,2},$\thanks{E-mail: m.nurisso@isac.cnr.it}
Annalisa Celotti$^{1,2,3,4}$\thanks{E-mail: celotti@sissa.it}, Andrea Mignone$^{5}$ and Gianluigi Bodo$^{6}$
\\
$^{1}$SISSA, Via Bonomea 265, I-34136 Trieste, Italy\\
$^{2}$INAF - Istituto Nazionale di Astrofisica, Osservatorio Astronomico di Brera, via Bianchi 46, I-23807 Merate, Italy\\
$^{3}$INFN - Istituto Nazionale di Fisica Nucleare, Sezione di Trieste, Via Valerio 2, I-34127 Trieste, Italy\\
$^{4}$IFPU - Institute for Fundamental Physics of the Universe, Via Beirut 2, I-34151 Trieste, Italy\\
$^{5}$Dipartimento di Fisica Generale, Universit\'a degli Studi di Torino , Via Pietro Giuria 1, I-10125 Torino, Italy\\
$^{6}$INAF - Istituto Nazionale di Astrofisica, Osservatorio Astrofisico di Torino, Strada Osservatorio 20, I-10025 Pino Torinese, Italy
}
\date{Accepted 2023 April 24; Revised: 2023 April 15; Received: 2022 August 17}

\pubyear{2022}

\begin{document}
\label{firstpage}
\pagerange{\pageref{firstpage}--\pageref{lastpage}}
\maketitle

\begin{abstract}
We present a new algorithm for the identification and physical characterization of current sheets and reconnection sites in 2D and 3D large scale relativistic magnetohydrodynamic numerical simulations. 
This has been implemented in the \textsc{PLUTO} code and tested in the cases of a single current sheet, a 2D jet and a 3D unstable plasma column. 
Its main features are: a) a computational cost which allows its use in large scale simulations; b) the capability to deal with complex 2D and 3D structures of the reconnection sites. 
In the performed simulations, we identify the computational cells that are part of a current sheet by a measure of the gradient of the magnetic field along different directions.
Lagrangian particles, which follow the fluid, are used to sample plasma parameters before entering the reconnection sites that form during the evolution of the different configurations considered.
Specifically, we track the distributions of the magnetization parameter $\sigma$ and the thermal to magnetic pressure ratio $\beta$ that - according to particle-in-cell simulation results - control the properties of particle acceleration in magnetic reconnection regions.
Despite the initial conditions of the simulations were not chosen \quotes{ad hoc}, the 3D simulation returns results suitable for efficient particle acceleration and realistic non-thermal particle distributions. 
\end{abstract}


\section{Introduction}\label{chap:intro}

Magnetic reconnection is a plasma process that dissipates the energy stored in magnetic field into plasma kinetic and thermal energy, through a rearrangement of magnetic field topology, resulting in particle heating and acceleration. 
It is thought to play an important role in several different astrophysical sources, including solar corona \citep[e.g.][]{Krucker_2013,Gary_2018,Pontin2022}, pulsar wind nebulae \citep[PWNe;][]{Kirk03,Cerutti_2014,Cerutti_2020}, Gamma-Ray Bursts \citep[GRBs;][]{Zhang2010,McKinney2011,Kumar_2015} and coronae and jets in Active Galactic Nuclei \citep[AGN;][]{Giannios2009,Nalewajko_2011,Sironi2015,Davelaar2020,Nishikawa_2020} \citep[See][for a review on recent progresses]{Guo_2020}. 

It is generally thought that pulsar winds and relativistic jets from GRBs and AGN are hydromagnetically launched \citep[see e.g.][]{Komissarov_2007,Komissarov_2009,Spruit2009,Chantry_2018}.
On the theoretical ground this appears to be the most effective and efficient way to generate highly relativistic outflows, as strongly supported also by numerical General Relativistic magnetohydrodynamics simulations \citep[][]{Tchekhovskoy2011,Tchekhovskoy2012,McKinney2012,Kaaz2022,Ripperda_2022}.
This process leads to the formation of a Poynting dominated jet and from an observational perspective the detected polarized non--thermal radiation supports the hypothesis that jets are highly magnetized at their base \citep[]{Doeleman2012, MartVidal2015}. 

The mechanism which leads to conversion from magnetic to  internal energy of the highly energetic particles required to emit the observed radiation has still to be understood. 
Diffusive shock acceleration (DSA) and magnetic reconnection (MR) are usually invoked as viable possibilities.
Shocks can be efficient in dissipating bulk kinetic energy but they are considered quite inefficient for highly magnetized plasma \citep{Sironi2015}. 
In such a situation MR could play a more relevant role. 
Both processes have been extensively studied both at the semi--analytical level \citep[for MR see][]{Drake_2010,Drake2012,Drury2012} as well as with particle--in--cell simulations (PIC) in which it is possible to describe self--consistently the interplay between magnetic fields and particles.

Specifically, PIC simulations have shown that MR in a magnetically dominated plasma effectively leads to particles acceleration, dissipating magnetic energy \citep[][]{Guo2014,Guo2015,Guo_2016,Jaroschek2004,Kilian2020,Li_2019,Lyutikov_2017,Petropoulou_2019,Sironi2014,Sironi16,Werner2015,Werner2017,Werner2017b,Werner_2021,Zenitani_2001,Zenitani2005,Zenitani_2008}.
Electrons (and protons) can be accelerated to ultra--relativistic energies, developing a high energy component, described as a power--law depending on the value of the magnetization $\sigma$, namely the ratio of magnetic to particle energy density (see below for details) \citep[]{Werner2015,Werner2017,Ball2018} and whose slope can be approximately estimated from the initial condition of the surrounding plasma. 
Further progress towards more realistic astrophysical scenarios, strives for restoring kinetic effects in MR in magnetohydrodynamics (MHD) simulations, in order to fill the gap between the micro-physical scale, at the skin--depth scale $c/ \omega_{\rm P}$ (where $\omega_{\rm P}$ is the plasma frequency and $c$ the speed of light) and the macroscopic one of astrophysical phenomena.

A first attempt in this direction is represented by MHD simulations with test particle distributions. 
Test particles are injected in the reconnection domain, allowing to study first and second--order Fermi acceleration due to MR \citep[][]{Kowal_2011,Kowal2012,Beresnyak2016,delValle2016,Medina_Torrej_n_2021,Puzzoni21}.
A further step to tackle the multi--scale challenge is to introduce sub--grid models in MHD codes, taking advantage of the results obtained with PIC simulations to describe the final spectra due to MR acceleration based on the condition of the plasma surrounding the reconnection region.

In a series of papers, we aim at studying particles acceleration as well as radiative emission signatures in large-scale numerical simulations through the introduction of a novel sub--grid model to capture localized magnetic reconnection regions.
This will be achieved using the hybrid particle--fluid framework of \citet[][]{Vaidya2018} available within the \textsc{PLUTO} code \citep[][]{Mignone2007}.
The non--thermal population is described by Lagrangian particles (macroparticles) representing an ensemble of real particles sufficiently close in physical space and described by their velocity and spectrum. 
Outside acceleration sites their spectra evolve according to the relativistic cosmic-ray transport equation, while MHD equations are solved concurrently. 
Their position is determined by a simple transport equation using the fluid velocity.
The characterization of the acceleration of macroparticles in reconnection sites follows a procedure similar to what is implemented in the \textsc{PLUTO} code, where a description of acceleration in diffusive shocks has been already introduced \citep{Vaidya2018,Mukherjee2021}.

In this initial part of the work, we present the first attempt - to the extent of our knowledge - to consistently include sub-grid MR models in large scale simulations, using predictions obtained by means of PIC numerical computations.
The sub--grid model is based on a current--sheet detection algorithm. 
Compared to previous methods \cite[e.g.][]{Servidio2010,Zhdankin2013,Kadowaki2018,Scepi2022}, we propose a more efficient, 3D--ready and time--dependent strategy to capture current sheet regions in large scale MHD simulations.

We compare such algorithm with the one proposed by \citet[][]{Zhdankin2013}, and sample the fluid quantities characterizing the dissipation process and the particle acceleration, namely the magnetization $\sigma$ and the ratio between plasma and magnetic pressures $\beta$, presenting their distributions. 
By using physically motivated prescriptions from PIC simulations, in a second paper (Nurisso et al, in preparation, hereafter Paper II) these quantities, associated with macroparticles that enter a current sheet, will be used to describe, for the first time in large scale simulations, non--thermal spectra of particles that are accelerated by MR and to model their non--thermal emission.

The paper is organized as follows: in Section \ref{chap:equations} we describe the relevant equations used in this paper; in Section \ref{chap:detector} we describe the current sheet detector and its validation; in Section \ref{chap:mr_char} we define the fluid quantities important to characterize the particles acceleration and describe how we sample them in MHD simulations. We apply the new numerical method to magnetized jet simulations in Section \ref{chap:jet_sim} and finally in Section \ref{chap:conclusions} we summarize the results and present our conclusions and future perspectives.
In the following the chosen units have $c = 1$.

\section{Relevant Equations}\label{chap:equations}

Our aim is the identification of current sheets in large-scale relativistic MHD (RMHD) simulations of a slab jet and a highly magnetized, relativistic plasma column.
The relevant equations are those of ideal relativistic MHD:

\begin{equation}\label{eq:RMHD_eqns}
\begin{aligned}
    &\partial_{\rm t} \left( \gamma \rho \right) + \nabla \cdot \left( \gamma \rho \mathbf{v} \right) = 0 \\
    &\partial_t \mathbf{m} + \nabla \cdot \left[ \gamma^2 w \mathbf{v}\mathbf{v} - \mathbf{E}\mathbf{E} - \mathbf{B}\mathbf{B} + (p + u_{\rm em}) \mathbf{I}\right] = 0 \\
    &\partial_{\rm t} \left( \gamma^2 w - p + u_{\rm em}  \right) + \nabla \cdot \left( \gamma^2 w \mathbf{v} + c \mathbf{E} \times \mathbf{B} \right) = 0 \\
    &\partial_{\rm t} \mathbf{B} + \nabla \times c \mathbf{E} = 0 \, ,
\end{aligned}
\end{equation}
where $\rho$ is the rest-mass density, $\mathbf{m} = \gamma^2 w \mathbf{v} + c \mathbf{E} \times \mathbf{B}$ the momentum density, $p$ the gas pressure, $w$ the relativistic enthalpy and $\gamma$ the Lorentz factor.
$\mathbf{E}$, $\mathbf{B}$ and $\mathbf{v}$ are, respectively, the three--vectors representing the electric field, the magnetic field and the velocity, $u_{\rm em} = (E^2 + B^2)/2$ and $\mathbf{I}$ is a $3 \times 3$ unit tensor.
The electric field is determined by the ideal condition $c \mathbf{E} + \mathbf{v} \times \mathbf{B} = 0$.
For a relativistic gas an accurate treatment would require the use of the Taub-Matthews equation of state (EoS), that is left for future improvements.
In the following an ideal EoS is adopted, so that  $w = \rho + \Gamma/(\Gamma - 1)p$ with a constant adiabatic index, here assumed $\Gamma = 5/3$ \citep[for a more general discussion on the relativistic EoS see][]{Mignone2007b}.
The units are chosen so that a factor $\sqrt{4 \pi}$ is reabsorbed in the definition of $\mathbf{E}$ and $\mathbf{B}$.

The system of Equations (\ref{eq:RMHD_eqns}) is solved by means of Godunov-type shock-capturing finite volume schemes, as illustrated in the \textsc{PLUTO} seminal paper by \cite{Mignone2007} (see also references therein).

\section{Current-sheet detector}\label{chap:detector}

Magnetic reconnection is thought to take place in current sheets, thin layers with large values of the current density, $\mathbf{J} = c / 4\pi \nabla \times \mathbf{B}$ (in the static case). 
Idealized current sheet configurations have been extensively adopted by PIC and MHD studies, typically in the form of a Harris current sheet, $\mathbf{B} = -B_{\rm 0} \mathrm{tanh} (y / \Delta) \mathbf{\hat{x}}$, where $B_{\rm 0}$ is the reconnecting field and $\Delta$ is the thickness of the current sheet.
However, when current sheets form dynamically as byproducts of plasma instabilities, we may expect more irregular shapes that can significantly differ from the idealized profile.

For these reasons we propose an algorithm to identify localized plasma region where reconnection may take place.
In particular, we aim at: i) improving the computational speed to identify current sheets at runtime during large-scale simulations; ii) tracing informations on the physical parameters (to be subsequently employed in particle spectral update) in complex structures. 
While other methods exist in literature, they are either complex to be extended to a 3D time-dependent simulation because based on the vector potential $\mathbf{A}$ \citep[][]{Servidio2010} or they rely on the definition of an average value of $\mathbf{J}$ on the computational domain \citep[e.g.][]{Zhdankin2013,Zhdankin2014,Zhdankin2015,Makwana2015,Kadowaki2018} that can be problematic when considering very inhomogeneous situations, such as that of a jet propagating into an external environment with very different properties.
Besides, these methods additionally identify a current sheet by clustering adjacent cells, part of the same acceleration site. 
This provides more informations about the global characteristic of the reconnection region, but it requires extra computational time.

The newly proposed algorithm follows from \citet[][]{Mignone2011} (see, in particular, Sec. 5.3 of their paper) and flags as reconnection sites cells that satisfy the following condition (in the 2D Cartesian case):

\begin{equation}
    \label{eq:cs_mignone2D}
    \chi = \frac{\left|\Delta_{\rm x} B_{\rm y}-\Delta_{\rm y} B_{\rm x}\right|}{\left|\Delta_{\rm x} B_{\rm y}\right|+\left|\Delta_{\rm y} B_{\rm x}\right| + c \sqrt{\rho}} > \chi_{\rm \min}\, ,  
\end{equation}
where $\Delta_{\rm x} B_{\rm y}$ and $\Delta_{\rm y} B_{\rm x}$ are undivided central differences and $\chi_{\rm min}$ is a free parameter.
Eq. \ref{eq:cs_mignone2D} represents an heuristic method to identify current sheets. 
The term $c\sqrt{\rho}$ has been included in order to avoid spurious artifacts in the presence of very low magnetic fields that could lead to divisions by zero.
The parameter $\chi$ gives a measure of the magnetic field gradient and it is  computed by performing finite differences on adjacent cells, making it computationally less expensive and more efficient in parallel computation with respect to the other method already cited. 
Its computation also does not need the definition of a region over which to average quantities are evaluated as in \citet[][]{Zhdankin2013}, leaving $\chi_{\rm min}$ as the only problem-dependent parameter.

Eq. (\ref{eq:cs_mignone2D}) can be extended to the 3D Cartesian case leading to the following expression for $\chi$:

\begin{equation}\label{eq:cs_mignone3D}
   \small
    \frac{\left|\Delta_{\rm x} B_{\rm y}-\Delta_{\rm y} B_{\rm x}\right| + 
          \left|\Delta_{\rm x} B_{\rm z}-\Delta_{\rm z} B_{\rm x}\right| +
          \left|\Delta_{\rm y} B_{\rm z}-\Delta_{\rm z} B_{\rm y}\right|}
    {\left|\Delta_{\rm x} B_{\rm y}\right|+\left|\Delta_{\rm y} B_{\rm x}\right| + 
     \left|\Delta_{\rm x} B_{\rm z}\right|+\left|\Delta_{\rm z} B_{\rm x}\right| + 
     \left|\Delta_{\rm y} B_{\rm z}\right|+\left|\Delta_{\rm z} B_{\rm y}\right| + c \sqrt{\rho}} \, .
\end{equation}

\subsection{Method comparison and verification}\label{chap:det_ver}

\begin{figure}
\centering
\includegraphics[width=\linewidth]{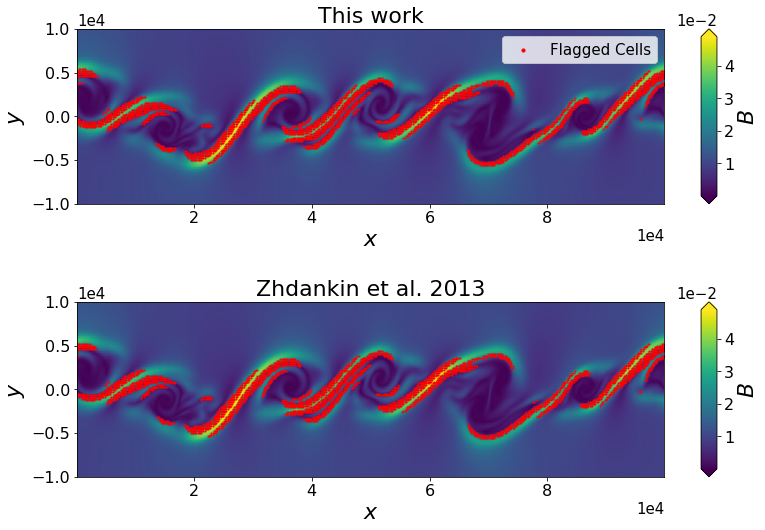}
\caption{Comparison of the results of the current sheet detection method described in Sec. \ref{chap:detector} (top panel) and that described in \citet{Zhdankin2013} (bottom panel) at $t = 6 \times 10^5 L/c$. The color plots show the magnetic field $B$ and the red dots represent points flagged as current sheets regions.}
\label{fig:comparison}
\end{figure}

In order to verify our method, we first compare our results with those obtained by adopting the method proposed by \cite{Zhdankin2013}, which identifies current sheet regions in MHD snapshots through local maxima of the current density $\mathbf{J}$. 
After the evaluation of the average value of the current density $\langle |\mathbf{J}|\rangle$, cells with $|\mathbf{J}|_{ijk} > n \langle |\mathbf{J}| \rangle$ are selected as candidate current sheet cells, where $n$ is a free parameter of the algorithm. 

To compare the two methods, we perform 2D MHD simulations of a Kelvin-Helmholtz instability (KHI) naturally producing current sheets.
The domain consists of a 2D rectangular box of size $L \times L/2$ with a resolution of $512 \times 256$ grid points.
The velocity has a profile $v_{\rm x} = 0.5 \, v_{\rm 0}\, {\rm sign}(y)$, where $v_{\rm 0} = 0.1 c$ is the shear velocity.
The instability is triggered by perturbing the $y$-component of velocity, $v_{\rm y} = r \, v_{\rm 0} \, {\exp} (-50 \, y /L_{\rm y})$, where $L_{\rm y}=L/2$ is the vertical size of the computational box and $r$ is a random number in the range $[-10^{-2},10^{-2}]$.
The initial density is $\rho_{\rm 0} = 1$ and the magnetic field is oriented along the $x$-direction with strength $B=0.1\sqrt{\rho c_s^2}$, where the sound speed $c_{\rm s}=v_{\rm 0} = 0.1 c$. 
The pressure is set as $p = c^2_{\rm s} \rho_{\rm 0} / \Gamma$, where $\Gamma = 5/3$ is the adiabatic index.
Boundary conditions are periodic on the $x$ direction and outflow elsewhere. 

The comparison between the two methods has been carried out in post-processing by varying $\chi_{\rm min}$ and $n$ and computing the number of flagged zones common to both algorithms.
The average value of $\mathbf{J}$, required by the method of \citet[][]{Zhdankin2013}, is evaluated over the whole domain.
Fig. \ref{fig:comparison} shows the results of the best agreement (maximum overlapping, with $\approx 96 \%$ of points found by the two algorithms), which is obtained at $t = 6 \times 10^5 L/c$ when $\chi_{\rm min} = 0.006$ and $n = 21.6$, supporting the effectiveness of our algorithm\footnote{Although Eq.\,\ref{eq:cs_mignone2D} is not rotationally invariant it is relevant to notice the agreement  in Fig. \ref{fig:comparison} comprises the presence of reconnection sites with different orientations.}. 
It has to be noticed that the average value of $\mathbf{J}$ over the whole domain greatly changes during the time evolution and the comparison with the method proposed by \citet[][]{Zhdankin2013} can be performed only for a specific time.
At a previous time $t = 5 \times 10^5 L/c$, for example, the maximum overlapping ($\approx 95 \%$) is achieved with $\chi_{\rm min} = 0.006$ and $n = 16$, due to the earlier development of the KHI.

\begin{figure*}
    \centering
    \includegraphics[width=\linewidth]{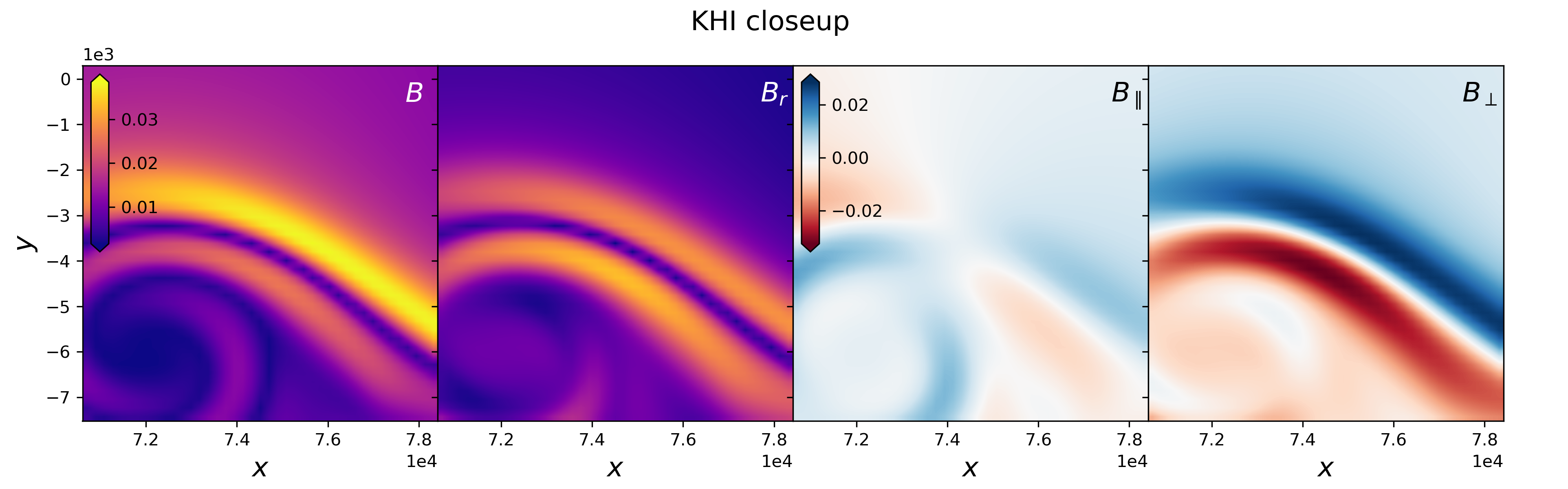}
    \caption{Total and reconnecting magnetic field intensities in the KHI case (zoom on a smaller
    domain patch). 
    From left to right: The $1^{\rm st}$ and $2^{\rm nd}$  panels represent, respectively, the  magnetic field $|\boldsymbol{B}|$ around a current sheet (identified via Eq. \ref{eq:cs_mignone2D}) and the reconnecting magnetic field $|\boldsymbol{B_{\rm r}}|$ after the subtraction of the mean field. 
    The $3^{\rm rd}$ and $4^{\rm th}$ panels show the parallel $|\boldsymbol{B_{\rm r}}|$ and perpendicular components to $\langle \nabla |\mathbf{J}|\rangle$, respectively.} 
    \label{fig:subtraction}
\end{figure*}

As done in \citet[][]{Zhdankin2013}, we want to check in post-processing that the current sheets found by the algorithm are regions in which an inversion of the magnetic field components parallel to the sheet is actually present. 
To this end, we estimate the direction normal to the current sheet by using $\nabla |\mathbf{J}|$ as a proxy, which we evaluate as an average over the cells surrounding the one in which $|\mathbf{J}|$ has a local maximum.
This allows us to have a reliable estimate of the vector perpendicular to the current sheet in its most dissipative regions.
We estimate a mean field $\mathbf{B}_{\rm m}$ as the average value of the magnetic field components in a box of $21 \times 21$ cells size. 
This region has been defined \quotes{ad hoc} for this problem in order to include the entire reconnection sites.
We then define a \quotes{reduced} field $\mathbf{B}_{\rm r}$ as $\mathbf{B}_{\rm r} = \mathbf{B} - \mathbf{B}_{\rm m}$. 
It is then possible to evaluate the parallel component $B_\parallel$ and the perpendicular one $B_\perp$ of $\mathbf{B}_{\rm r}$ with respect to $\nabla |\mathbf{J}|$.
In Fig. \ref{fig:subtraction} we show the result for a single current sheet found in the simulation. 
The $1^{\rm st}$ and $2^{\rm nd}$ panels show, respectively, the total magnetic field intensity $\mathbf{B}$ and the reduced magnetic field $\mathbf{B}_{\rm r}$, that can be considered a good approximation of the reconnecting field after the mean field subtraction. 
The $3^{\rm rd}$ and $4^{\rm th}$ panels show $B_\parallel$ and $B_\perp$, i.e. the parallel and perpendicular components of $\mathbf{B}_{\rm r}$ with respect to the current sheet perpendicular $\nabla |\mathbf{J}|$.
Notice that $\mathbf{B}_{\rm r}$ results more symmetric along the current sheet direction with respect to the total magnetic field $\mathbf{B}$ and that $B_\perp$ shows the inversion of polarity expected for a current sheet in which magnetic reconnection takes place.

\section{Characterization of Magnetic Reconnection properties}\label{chap:mr_char}

In order to estimate the efficiency of particle acceleration 
and the particle spectra resulting from our simulations, it is necessary to determine the physical properties of the identified reconnection sites.
PIC studies \citep{Werner2017, Ball2018} have shown that the plasma quantities that play a major role in this respect are the cold ion (proton) magnetization $\sigma$ and $\beta$ (the ratio between proton thermal pressure and magnetic pressure) defined as:

\begin{equation}\label{eq:sigma_beta}
    \sigma \equiv \sigma_{\rm i} \equiv \frac{B^{2}}{4 \pi n_{\rm i} m_{\rm i} c^2}
    \qquad
    \beta \equiv \beta_{\rm i} \equiv \frac{8 \pi n_{\rm i} k T_{\rm i}}{B^2} \, ,
\end{equation}

\noindent where $B$ refers to the magnetic field that undergoes reconnection, $n_{\rm i}$ is the ion density and $T_{\rm i}$ their temperature.
These quantities refer to the plasma region surrounding the reconnection site.
In the following we will use the symbols $\sigma$ and $\beta$ to refer to the fluid quantities of Eq. \ref{eq:sigma_beta}.

The particle spectrum resulting from MR has been indeed extensively studied in a fully kinetic framework with PIC simulations, both in a pair plasma \citep[]{Zenitani_2001,Jaroschek2004,Zenitani2005,Zenitani_2008,Guo2014, Sironi2014, Guo2015,Werner2015,Sironi16,Werner2017b,Petropoulou2018,Werner_2021,Zhang_2021}, in a ion-electron \citep{Melzani2014,Guo_2016,Werner2017,Ball2018,Li_2019,Kilian2020} and in a pair-ion one \citep[][]{Petropoulou_2019}. 

The spectrum is (at $1^{\rm st}$-order) described as a quasi--power--law distribution in particle energy, $f(\epsilon) \propto \epsilon^{-p}$, where $\epsilon = (\gamma_{\rm p} -1)m c^2$ and  $\gamma_{\rm p}$ is the Lorentz factor of the particles represented by a macroparticle, with a high-energy cut--off. 
In particular in \cite{Werner2017} the particles spectrum has been studied in the semi-relativistic regime $(10^{-3} < \sigma < 1)$ and up to the relativistic one $(\sigma \gg 1)$. 
Their work has shown that the expected power--law index $p$ and cut--off energy can be approximated as a function of $\sigma$, in agreement with the results by \cite{Ball2018}. 
While this holds for $\beta \lesssim 3\times 10^{-3}$, \cite{Ball2018} found that at higher values the power--law steepens and the final index $p$ depends also on the value of $\beta$. 
The other two parameters needed for the description of the post--reconnection spectrum, i.e. the fraction of energy gained by the electrons and the acceleration efficiency, also depend on the same fluid parameters $\sigma$ and $\beta$ \citep[][]{Werner2017,Ball2018}.
With the sampling of these two quantities it is thus possible to reasonably approximate the particle spectra.
In Paper II we will focus on the results for an ion--electron plasma and the electron acceleration through the use of macroparticles, as described below, and a sub--grid model based on PIC simulations results. 

The values usually assumed in PIC studies (in order to determine the initial configuration of the current sheet) are the asymptotic values of $\sigma$ and $\beta$, far from the reconnecting region. 
In what follows, the corresponding $\sigma$ and $\beta$ will be identified with the respective fluid quantities around the region recognized as current sheet in the MHD simulations.
In the presence of complex structures and magnetic field configurations evolving with time, our current sheet identification only flags cells where acceleration is supposed to take place (but it does not reconstruct the entire reconnection region).
Therefore a more sophisticated procedure with dynamic sampling is needed.
To this purpose, we develop an algorithm that makes use of macroparticles, comoving with the fluid, which sample $\sigma$ and $\beta$ at simulation runtime and keep track of them.

The algorithm is implemented in the \textsc{Lagrangian Particle} module \citep{Vaidya2018} in the \textsc{PLUTO} code \citep{Mignone2007}. 
In this module the spatial motion of macroparticles is described by:

\begin{equation}\label{eq:macroparticles_motion}
    \frac{d \textbf{x}_{\rm p}}{dt} = \textbf{v}(\textbf{x}_{\rm p})\, ,
\end{equation}
where $\textbf{v}$ represents the fluid velocity interpolated at the macroparticle's position and the subscript \quotes{p} labels the macroparticle.

While the macroparticles move in the domain according to Eq. (\ref{eq:macroparticles_motion}), $\sigma$ and $\beta$ at the particle's position are sampled at each step and their values are stored in the variables $\sigma_{\rm p}$ and $\beta_{\rm p}$.
In order to ensure optimal sampling of $\sigma$ and $\beta$ as the particles move through the reconnection region, we introduce an algorithm that considers the history of $\sigma_{\rm p}$ and $\beta_{\rm p}$.
This allows us to correctly employ parameters from PIC studies, which may not correspond to values sampled in the last cell before entering the reconnection region, nor to the extreme values encountered by the macroparticles in their trajectories.
In what follows we give a description of the algorithm:

\begin{enumerate}
    \item As soon as a each macroparticle exits a reconnection region or it is injected in the simulation domain, the corresponding values of $\sigma$ and $\beta$ are sampled by interpolating the fluid quantities at the macroparticle's position and stored in the variables $\sigma_{\rm p}$ and $\beta_{\rm p}$. 
    The variable $N$, describing the number of steps from the reset of $\sigma_{\rm p}$ and $\beta_{\rm p}$ values, is set to $N=1$.
    \item At each time step, while the macroparticle is in a cell that has not been tagged as current sheet region, the stored values are updated with the sampled ones only if a new peak of $\sigma$ is detected.
    \item In case $\sigma$ does not represent a new peak, the stored values are updated through a weighted average, namely:
    
    \begin{equation}
    \begin{aligned}
        \sigma_{\rm p} &\leftarrow \sigma_{\rm  {p,N-1}} + \frac{1}{\rm N} \left( \sigma - \sigma_{\rm {p,N-1}} \right) \\
        \beta_{\rm p} &\leftarrow \beta_{\rm {p,N-1}} + \frac{1}{\rm N} \left( \beta - \beta_{\rm {p,N-1}} \right)\, ,
    \end{aligned}
    \end{equation}
    
    \noindent where $\sigma_{\rm {p,N-1}}$ and $\beta_{\rm {p,N-1}}$ represent the values previously associated to the macroparticle and $N$ is the number of steps from the last reset of the sampled quantities. 
    In this way, $\sigma_{\rm p}$ and $\beta_{\rm p}$ represent averages over the last portion of the macroparticle trajectory and progressively forget previous maxima as the macroparticle moves away from them.
    \item If the macroparticle lies in a tagged cell, the values of $\sigma_{\rm p}$ and $\beta_{\rm p}$ are not updated and are taken as representative of the asymptotic values with which the macroparticle has entered the current sheet.
\end{enumerate}

With the average of the sampled $\sigma$ and $\beta$ 
we ensure that $\sigma_{\rm p}$ and $\beta_{\rm p}$ remain good estimates of the values surrounding the current sheets, independently of possible peculiar behaviours of the plasma  during its evolution far away from the reconnection sites.

\subsection{Analytical Current sheet: Numerical set-up}\label{chap:mr_char_set}

\begin{figure}
    \centering
    \includegraphics[width=\linewidth]{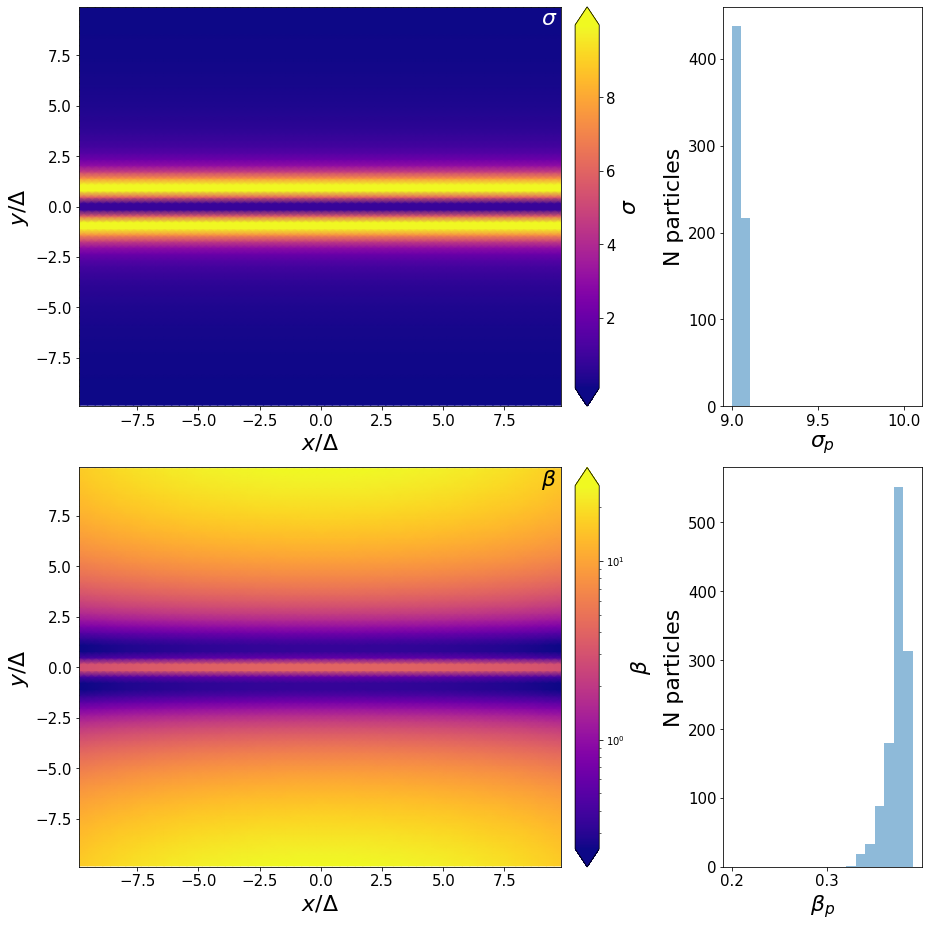}
    \caption{Results of the sampling of $\sigma_{\rm p}$ and $\beta_{\rm p}$ for the analytical set-up
    (\S\ref{chap:mr_char_set}). 
    The top and bottom left-hand panels represent, respectively, $\sigma$ and $\beta$ (in logarithmic scale) at $t = 30 \Delta / c$. 
    The right-hand panels show the corresponding distributions of $\sigma_{\rm p}$ and $\beta_{\rm p}$ sampled by the macroparticles located inside the reconnection region at that time.}
    \label{fig:steady_rec}
\end{figure}

We test the algorithm by determining $\sigma_{\rm p}$ and $\beta_{\rm p}$ in a 2D simulation with a set-up representing a current sheet of known values of $\sigma$ and $\beta$ of the fluid. 
Following \citet[][]{Chiuderi2014} (see their Section 9.1) we describe a stationary reconnection state, with velocity field described by:

\begin{equation}
    \mathbf{v} \equiv \left[\frac{v_0}{a} x, -\frac{v_0}{a} y, 0 \right]\, ,
\end{equation}
where $v_0$ and $a$ are constant values.
The electric field is directed along the $z$ direction, $\mathbf{E} = E \hvec{e}_z$, where $E$ is the constant field strength.
The magnetic field is assumed to lie in the $x$-direction, that is $\mathbf{B} \equiv \left[ B(y),0,0 \right]$ where the exact form of $B(y)$ can be recovered from the stationary condition ($\partial/\partial_t = 0$) and the Ohm's law for a resistive plasma.
This yields

\begin{equation}
    B(y) = B_0\, {\rm e}^{-(y/\Delta)^2} \int_0^{y/\Delta} {\rm e}^{u^2} {\rm d}u\, , 
\end{equation}
\noindent where $B_0 = 2E c a / v_0 \Delta$, $\Delta = \sqrt{2 \eta a / v_0}$ and $\eta$ is the magnetic diffusivity.

We assume a pressure profile of the form

\begin{equation}\label{eq:chiuderi_prs}
    p = c_0 + f(y) + c_x x^2 + c_y y^2 \,.
\end{equation}

In order to have a stationary solution we obtain the following conditions

\begin{equation}
    c_x = c_y = -\frac{\rho v_0^2}{2 a^2} \qquad f(y) = -\frac{1}{8 \pi} B^2(y) \quad ,
\end{equation}
while the choice of $c_0$ has to guarantee that $p>0$ over the whole domain.
Notice that the pressure profile contains a component proportional to $x^2$ and $y^2$, introducing a dependence on $\beta$, which is a function of the horizontal distance from the centre of the simulation.
This dependence is expected to be reflected in the distribution of the $\beta_{\rm p}$ sampled by the macroparticles.

This set-up, with the $B(y)$ and $p(y)$ profiles showing respectively a peak and a minimum for $y/\Delta = \pm 1$, is particularly suited to verify our sampling algorithm. In fact, the macroparticles with initial position $|y/\Delta|>1$  start with $\sigma_{\rm p}$ and $\beta_{\rm p}$ values that are reset when they encounter the peak of $\sigma$ at $y/\Delta = \pm 1$, allowing us to verify both the condition of reset of the algorithm and the agreement between the sampled $\sigma_{\rm p}$ and $\beta_{\rm p}$ and the fluid ones as defined in Eq. (\ref{eq:sigma_beta}).

The set-up consists of a 2D square domain of size $L \times L$, with $x / \Delta, y / \Delta \in \left[ -10,10 \right]$.
The resolution is kept deliberately low ($64 \times 64$ cells), in order to have the resolution of only a few cells on the vertical of the current sheet, similarly to what expected  in the simulations of Sec. $\ref{chap:jet_sim}$.
The boundary conditions are outflow everywhere. We set $v_0/a = 0.1$, $\eta = 0.05$, and $\rho =10^{-2}$ to be constant over the whole domain, $\sigma_{\rm peak} \equiv \sigma(y/\Delta = \pm 1) = 10$.
Macroparticles are injected at $t=0$ in the region $|y/ \Delta|>3$, so that their initial sampled value is far from the reconnection region and from the expected maximum values of $\sigma$ and $\beta$. 
The fluid is kept frozen in the initial configuration, while the macroparticles evolve according to Eq. (\ref{eq:macroparticles_motion}). 
We set $\chi_{\rm min} = 0.4$ as the threshold to identify the reconnection region, defined in Eq. (\ref{eq:cs_mignone2D}).

\subsubsection{Results on MR characterization}\label{chap:mr_char_res}

The results of the sampling of $\sigma_{\rm p}$ and $\beta_{\rm p}$ are shown in Fig. \ref{fig:steady_rec}. 
The top and bottom panels on the left--hand side represent the plasma values of $\sigma$ and $\beta$ (in logarithmic scale), respectively.
As expected, $\sigma$ peaks in a region along the entire $x-$axis at $y / \Delta = \pm 1$, with $\sigma_{\rm max} = 10$. 
A similar profile is found for $\beta$, where at $y / \Delta = \pm 1$ a minimum of $\beta_{\rm min} \sim 0.34$ is found.
The right--hand panels show the results of the sampling of $\sigma_{\rm p}$ and $\beta_{\rm p}$ for the macroparticles that are inside the reconnection region at time $t = 30 \Delta / c$.
The algorithm correctly resets their values at $y / \Delta = \pm 1$. The distributions found for $\sigma_{\rm p}$ and $\beta_{\rm p}$ are narrowly peaked at values close to the fluid $\sigma_{\rm max}$ and $\beta_{\rm min}$. More precisely, the sampled $\sigma_{\rm p}$ are systematically lower with respect to the peak value: this can be ascribed to the number of averaging operations ($N \sim 9-10$ in this set-up) that the sampling method performs before the macroparticles enter the reconnection region. The larger spread of the distribution of $\beta_{\rm p}$  
with respect to that of $\sigma_{\rm p}$ is due to the pressure profile, as mentioned above.

\begin{figure}
    \centering
    \includegraphics[width=\linewidth]{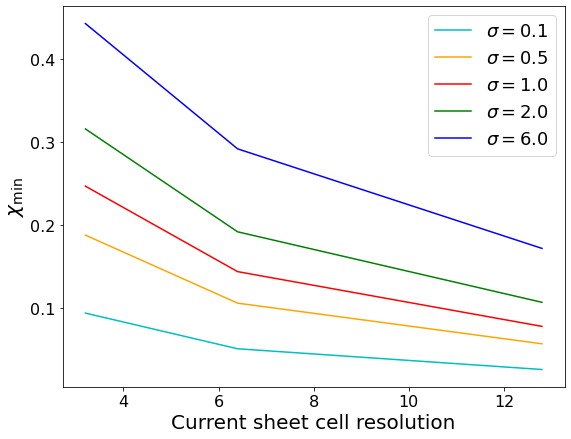}
    \caption{Values of the threshold $\chi_{\rm min}$ (Eq. \ref{eq:cs_mignone2D}) required to sample different values of $\sigma$ as a function of the resolution, given by the number of zones contained in the current sheet width.
    Different colours refer to different values of $\sigma$, as indicated.}
    \label{fig:res_sigma}
\end{figure}

We want to study also the dependence of the threshold $\chi_{\rm min}$ on the numerical resolution of a current sheet and on the values of $\sigma$ which can be detected, and thus sampled, by the algorithm (see Eq. \ref{eq:cs_mignone2D}). 
$\chi_{\rm min}$ has been determined as $\chi_{\min} \equiv \chi \left( y/\Delta = \pm 0.5 \right)$.
The resolution is indicated by the number of cells across the current sheet. 
In the following simulations, presented in Sec. \ref{chap:2Djet} and \ref{chap:3Djet},
we indicatively expect that the resolution, constrained by the computational cost, corresponds to a few cells in a single current sheet.

Fig. \ref{fig:res_sigma} shows the results. 
Clearly, for a given number of cells, lower values of the threshold are required to sample lower values of $\sigma$.
In the perspective of particle acceleration, the criteria based on Eq. (\ref{eq:cs_mignone2D}) favour reconnection regions with higher values of $\sigma$, indeed the more efficient ones for the acceleration of particles.  
In fact, the value of $\chi$ decreases with  $\sigma$ and therefore, for a given threshold,  regions with decreasing $\sigma$ will be progressively more excluded.
For a given $\sigma$, an increase in the number of cells requires a lower $\chi_{\rm min}$ for the algorithm to detect a reconnection site (by the definition of $\chi$).

While these results cannot be simply extended to more complex situations, they can still be indicative for the choice of the threshold for a given resolution and \quotes{required} minimum $\sigma$ to be sampled. 
 
\section{Jet simulations}\label{chap:jet_sim}

We now wish to assess the validity of our method on more complex configurations like those typically found in magnetically dominated jets.

\subsection{Slab Jet}\label{chap:2Djet}

\subsubsection{Numerical set-up}\label{chap:2Djet_setup}

The set-up consists of a 2D rectangular domain of size $L \times L/2$, with a resolution of $2048 \times 1024$ grid points.  
The ambient medium has constant pressure $p_{\rm 0} = 2 \times 10^{-3}$, density $\rho_{\rm 0} = 1$ and a magnetic field along the $x$--axis with constant magnitude $B_{\rm 0} = \sqrt{2 p_{\rm 0} /\beta_0}$, where $\beta_0$ is the plasma beta.
The jet enters the domain from a nozzle of radius $r_{\rm j} = 1$ along the $x$--direction, with speed $v_{\rm j} = 0.95 c$.
The box size can be expressed as function of $r_{\rm j}$ having $L = 40 r_{\rm j}$.
The jet pressure is assumed to be the same as that of the ambient medium, $p_{\rm j} = p_{\rm 0}$, while its density is $\rho_{\rm j} = \lambda \rho_{\rm 0}$, where $\lambda = 10^{-2}$ represents the density ratio.
Both quantities are expressed in the fluid rest frame.
With the chosen resolution, the initial slab jet is resolved with $\approx 50$ cells.
The macroparticles are injected at a fixed time interval $\Delta t_{\rm inj} = 1$ at the jet base, in a region $x \in \left[0,1\right]$ and $y \in \left[ -r_{\rm j},r_{\rm j} \right]$ with one macro-particle per cell.
The boundary conditions are outflow everywhere except for the injection region.

Values of $p$ and $\rho$ in the ambient medium have be chosen so that the ensuing jet is not ballistic.
We consider different values of $\beta_0$ in order to determine the expected sampled values of $\sigma_{\rm p}$ and $\beta_{\rm p}$. 
Specifically, we set the initial $\beta_0$ to be equal to $\beta_0 = 1/5$ and $\beta_0 = 1/15$, corresponding to $\sigma = 2$ and $\sigma = 6$.

We run the simulations with different values of $\chi_{\rm min} = 0.3, 0.2, 0.1$ and for $\chi_{\rm min} =0.1$ also at a lower resolution of $1024 \times 512$ grid points.

\subsubsection{Results}

\begin{figure*}
    \centering
    \includegraphics[width=\linewidth]{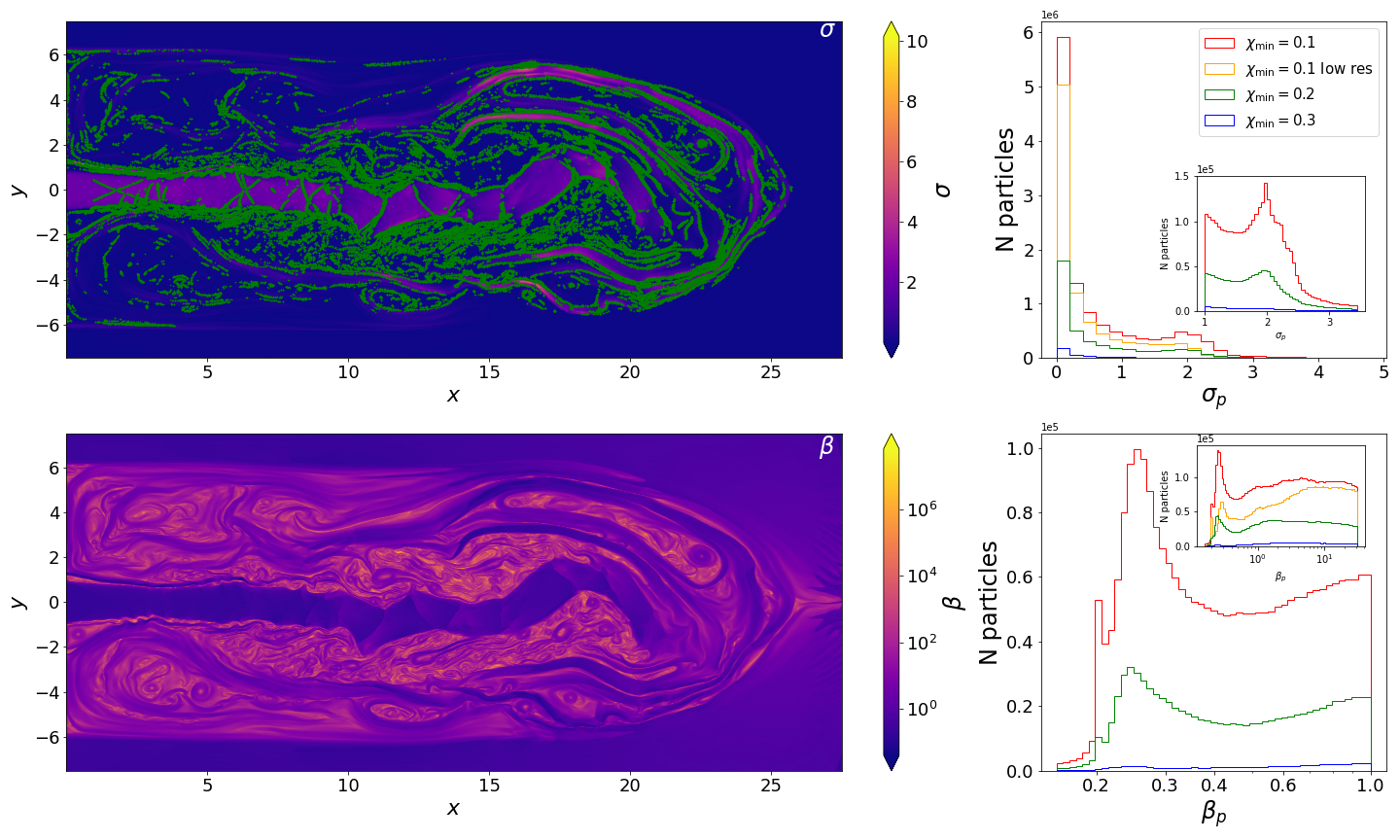}
    \caption{Results for the 2D MHD jet simulation with continuous injection, initial $\sigma = 2$, $\chi_{\rm min} = 0.1$ at the higher resolution.
    The left-hand panels show $\sigma$ and $\beta$ (in logarithmic scale) of the fluid at $t=112 r_{\rm j}/c$. 
    In the upper left panel the macroparticles lying inside a reconnection region at $t =112 r_{\rm j}/c$ are marked as green dots. 
    The right-hand panels represent the distributions of $\sigma_{\rm p}$ and $\beta_{\rm p}$ sampled by the macroparticles that entered a reconnection region in the time interval $t = (112 \pm 10) r_{\rm j}/c$. 
    The distributions for different thresholds $\chi_{\rm min}$ and resolutions are also plotted. 
    For the sake of clarity the histogram of $\sigma_{\rm p}$ around $\sigma_{\rm p} \sim 2$ is shown in the inset of the $\sigma_{\rm p}$ distribution. 
    For $\beta_{\rm p}$ the main distribution plot is limited to $\beta_{\rm p}<1$, while the whole distribution is presented in the inset. 
    }
    \label{fig:sigma_beta_histo_2Dsigma5}
\end{figure*}

\begin{figure*}
    \centering
    \includegraphics[width=\linewidth]{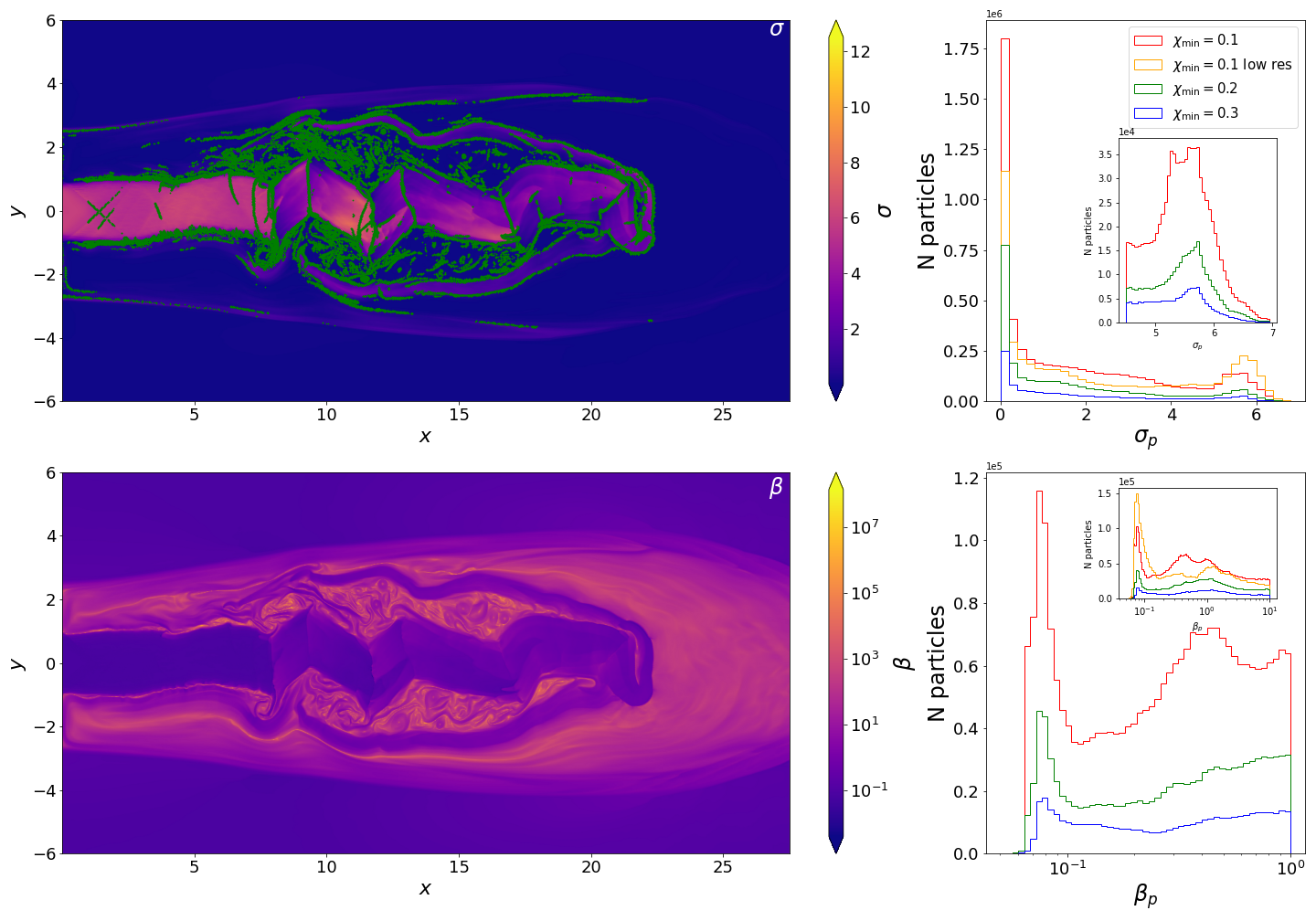}
    \caption{
    Same as Fig. \ref{fig:sigma_beta_histo_2Dsigma5} but for $\sigma = 6$.
    For the sake of clarity the histogram of $\sigma_{\rm p}$ around $\sigma_{\rm p} \sim 6$ is plotted in the inset of the $\sigma_{\rm p}$ distribution. 
    For $\beta_{\rm p}$ the main distribution plot is limited to $\beta_{\rm p}<1$, while the whole distribution is shown in the inset.
    }
    \label{fig:sigma_beta_histo_2Dsigma15}
\end{figure*}

The results obtained from the simulation with $\sigma = 2$ are shown in Fig. \ref{fig:sigma_beta_histo_2Dsigma5} at $t=112 r_{\rm j}/c$ (corresponding to $\approx 3$ light--crossing times of the entire domain).
The jet, injected from the left-side as described in Sec. \ref{chap:2Djet_setup}, interacts with the ambient medium.
The magnetic field in the external part of the jet, initially aligned along the $x-$direction, is distorted while interacting and forms current sheets.
The core of the jet instead remains more stable and less reconnection regions are observed.
At $t=112 r_{\rm j}/c$ the magnetic field is stretched, with macro--structures that are interacting with the ambient medium, visible as the regions with the highest values of $\sigma$.
In the top and bottom left--hand panels the values of $\sigma$ and $\beta$ (defined in Eq. \ref{eq:sigma_beta}) are plotted, respectively (in logarithmic scale for $\beta$). 
As expected, nearby the injection region, their values remain similar to the initial ones while, when interacting with the external medium, the values of $\sigma$ ($\beta$) tend to decrease (increase).
The distributions of the corresponding quantities $\sigma_{\rm p}$ and $\beta_{\rm p}$ sampled by the macroparticles that entered a reconnection region at times $t = (112 \pm 10) r_{\rm j}/c$ are shown in the right-hand panels.
The different histograms refer to different threshold values ($\chi_{\rm min}$) and grid resolutions.
The corresponding macroparticles positions are marked in the top left--hand panel using green dots.
These macroparticles are inside a reconnection region at $t = 112 r_{\rm j}/c$.

The $\sigma_{\rm p}$ distribution  is not monotonic: most of the reconnection sites have very low values of $\sigma_{\rm p}$, but another peak is observed at $\sigma_{\rm p} \approx 2$. 
The distribution of $\sigma_{\rm p}$ around this peak is shown in the inset of the histogram. 
Such a behaviour reflects the fact that the sampling macroparticles can enter reconnection sites lying in strongly magnetized regions (near the jet beam) as well as in the cocoon.
The values of $\beta$ span a broad range with only a small fraction achieving $\beta \lesssim 1$. 
The distribution of $\beta_{\rm p}$ has two peaks (around $\beta_{\rm p} \approx 0.3$ and a broad one around $\sim 10$) in correspondence of the two maxima of $\sigma_{\rm p}$. 
For the sake of clarity the main plot shows the distribution for $\beta_{\rm p}<1$, while the whole range spanned is plotted in the inset. 

Fig. \ref{fig:sigma_beta_histo_2Dsigma15} reports the same quantities for the case with $\sigma = 6$: consistently, the second peak of the $\sigma_{\rm p}$ distribution for this case is located around $\sigma_{\rm p} \approx 6$, in agreement with the value injected at the base of the jet. 
These results confirm that our method can correctly sample the reconnection sites as well as the values of magnetization and plasma $\beta$.

Regarding the choice of $\chi_{\min}$ (which determines the total number of sampled sites), we observe the peaks in the distributions start to become significant for $\chi_{\rm min} \sim 0.2$ and are better sampled for $\chi_{\rm min}= 0.1$. 
This is even more clear for the set-up with initial  $\sigma = 6$ due to the dependence of the value of $\chi_{\rm min}$ on the value of the minimum $\sigma_{\rm p}$ that the algorithm can sample (see Sec. \ref{chap:mr_char_res}). 

Finally, in both simulations we compare the results obtained with a lower resolution: no dramatic differences are found, with some differences in the number of sampling particles and distribution of $\sigma_{\rm p}$ and $\beta_{\rm p}$ that can be ascribed to the different simulations that have been performed to obtain these results. 

\begin{figure}
    \centering
    \includegraphics[width=\linewidth]{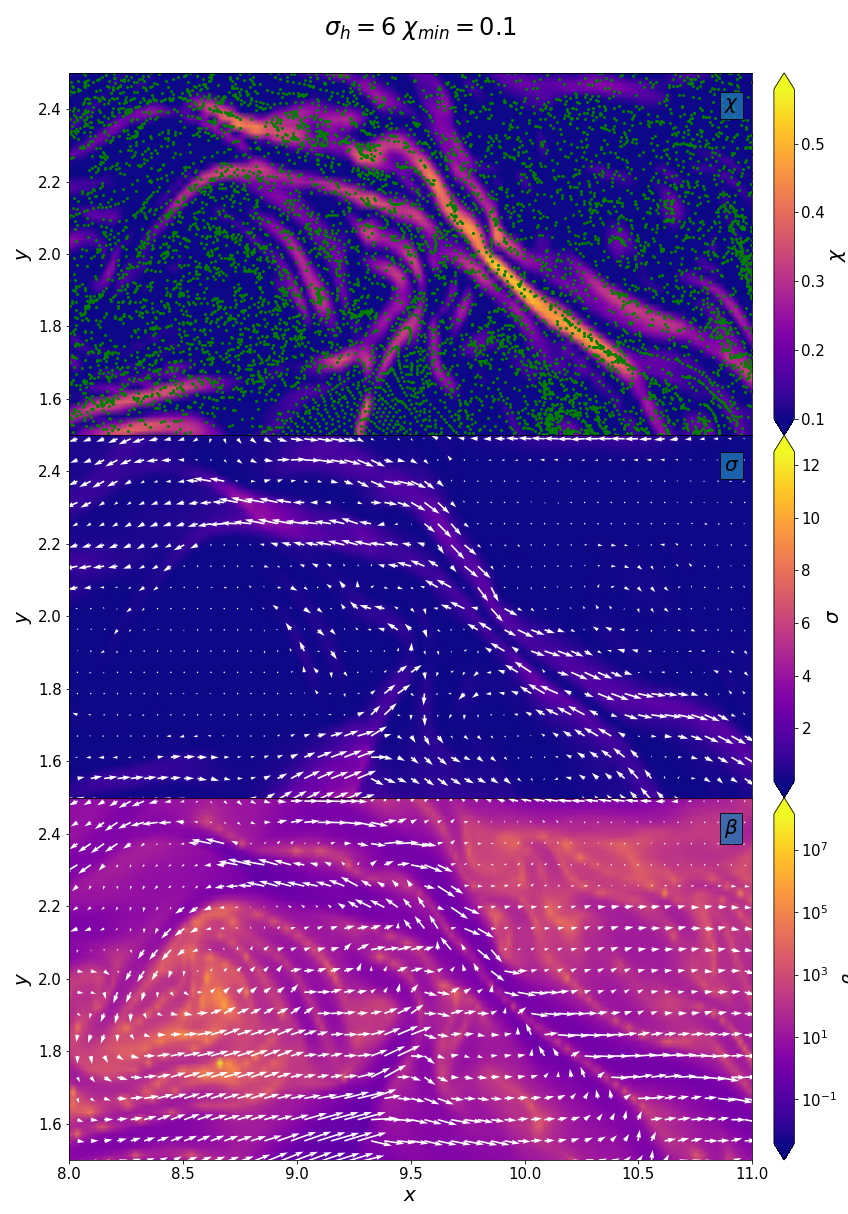}
    \caption{Closeup view of the region $x \in \left[8,11 \right]$ and $y \in \left[15, 2.5 \right]$ at $t=112 r_{\rm j}/c$ for the 2D RMHD jet simulation, shown in Fig. \ref{fig:sigma_beta_histo_2Dsigma15}.
    \textbf{Upper panel:} colormap of $\chi$ used to identify reconnection regions (Eq. \ref{eq:cs_mignone2D}). The minimum value has been set to the chosen threshold $\chi_{\rm min} = 0.1$. 
    Macroparticles located inside the identified sites are marked as green points. 
    \textbf{Central and bottom panels:} colormaps of $\sigma$ and $\beta$ with arrows representing the magnetic field (central) and velocity field (bottom).
    The length of the arrows is proportional to the magnitude.}
    \label{fig:2Dsigma15_zoom}
\end{figure}

For a better understanding of the behaviours of field and macroparticles in these large scale set-ups, a zoom on a reconnection region of the simulation of Fig. \ref{fig:sigma_beta_histo_2Dsigma15}, around $x \approx 10$ and $y \approx 2$, at $t=112 r_{\rm j}/c$ is reported in Fig. \ref{fig:2Dsigma15_zoom}. 
The region at the center of the plot is identified by the algorithm as a reconnection site, with many other smaller current sheets around it. 
The top panel shows the values of $\chi$ (for $\chi \geq 0.1$) together with the positions of the macroparticles, represented as green dots. 
Many macroparticles are inside this reconnection site at the time of the snapshot, after they sampled $\sigma_{\rm p}$ and $\beta_{\rm p}$ while entering it. 
The values of $\sigma$ are plotted in the central panel together with the magnetic field topology indicated by arrows whose length is proportional to the field strength. 
Around the reconnection site, $\sigma$ is higher with respect to its center, with $\sigma \gtrsim 1$. 
Its values are not symmetric on the two sides of the current sheet, reflecting the asymmetry of the configuration. 
Asymmetric configurations with different values of $\sigma$ around the reconnection site have also been recently investigated with PIC simulations by \citet[][]{Mbarek22} showing that relativitic asymmetric reconnection still produce power-law distributions and the slope depends on the magnetization of both inflowing plasmas.
The same behaviour can be taken into account by our method, with macroparticles entering from both sides.

Notice also that the magnetic field lines show inversion of the direction along the perpendicular to the current sheet, as we expect in reconnection events.

In the bottom panel, the values of $\beta$ are presented together with the velocity field of the particles. 
Although the values of $\sigma$ for this current sheet are typically large, most of the plasma is moderately magnetized ($\beta \gtrsim 1$) and only the sides of the most powerful reconnection region in the figure have $\beta \lesssim 1$.
The velocity field follows a behaviour similar to the magnetic field lines, as we expect in ideal RMHD.

\begin{figure}
    \centering
    \includegraphics[width=\linewidth]{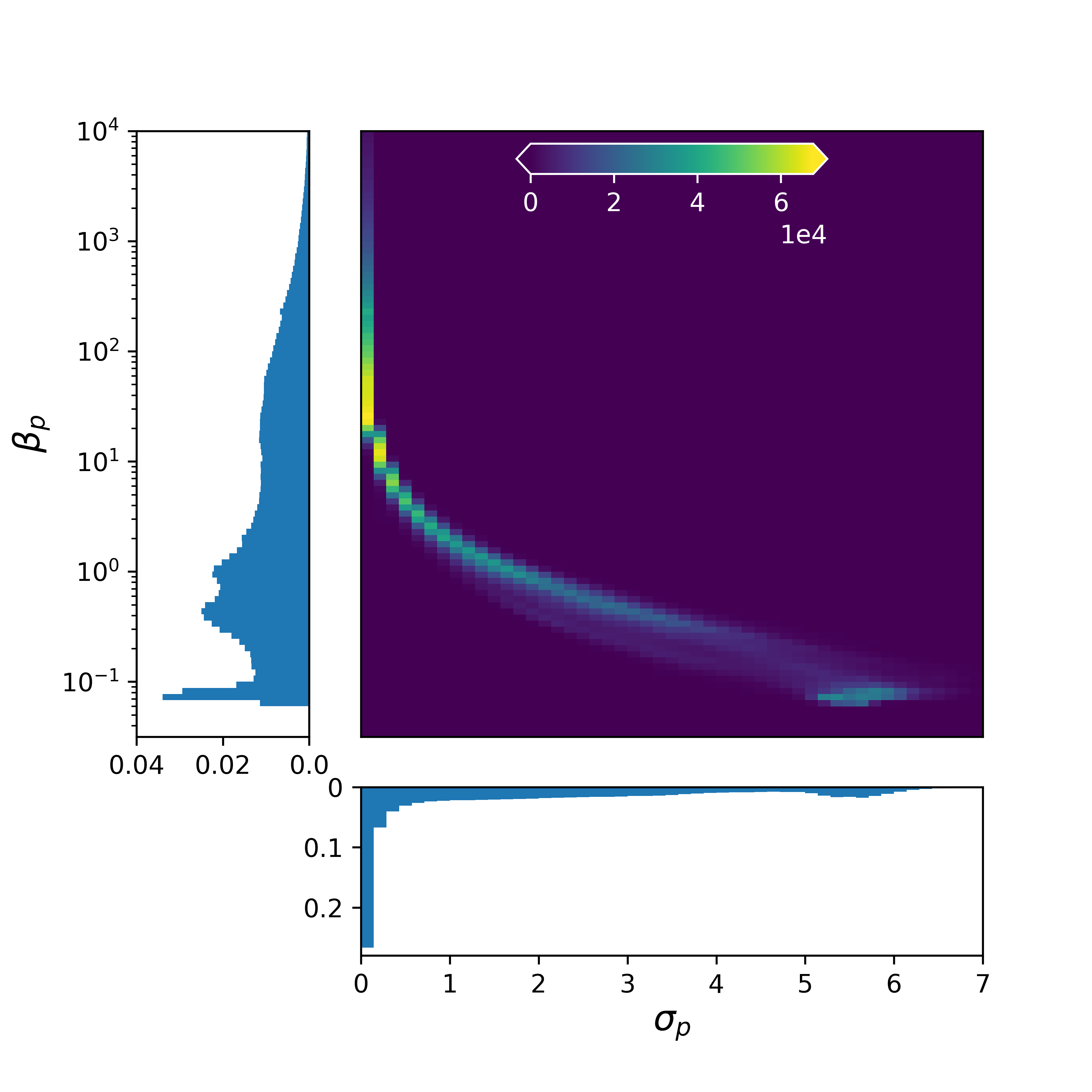}
    \caption{Values of $\sigma_{\rm p}$ and $\beta_{\rm p}$ for the 2D RMHD jet simulation with injected $\sigma = 6$ at $t = (112 \pm 10) r_{\rm j}/c$ (corresponding to Fig. \ref{fig:sigma_beta_histo_2Dsigma15}). 
    The left and bottom plots show the 1D histograms of the probability distribution functions, while the combined probability distribution is represented in the 2D plot, where the colour corresponds to the number of particles. 
    The probabilities represented refer to the whole domain.}
    \label{fig:sigma_vs_beta}
\end{figure}

Although we will amply discuss the following point in Paper II, it is worth noticing that the majority of the reconnection sites that these set-ups generate have values of $\beta_{\rm p}$ that are too large for efficient acceleration of particles to relativistic energies: in this case the dissipated energy may increase the temperature of the fluid. 
This can be more clearly seen in Fig. \ref{fig:sigma_vs_beta}, a 2D histogram of the values of $\sigma_{\rm p}$ and $\beta_{\rm p}$ for the case with injected $\sigma = 6$ (corresponding to Fig. \ref{fig:sigma_beta_histo_2Dsigma15} and $\chi_{\rm min} =0.1$). 
The color bar in the 2D histogram describes the number of particles with specific $\sigma_{\rm p}$ and $\beta_{\rm p}$ values. 
The corresponding 1D histograms show their probability distributions function.
Most of the sites with the lowest values of $\sigma_{\rm p}$ are associated to the tail of the $\beta_{\rm p}$ distribution at high values. 
A population of particles with favourable acceleration condition is present at high values of $\sigma_{\rm p}$ and  $\beta_{\rm p} \simeq 10^{-1}$. 
We stress however that in this configuration these values depend on the fluid properties set for the simulations.

\subsection{3D unstable plasma column}\label{chap:3Djet}

\subsubsection{Numerical set-up}

We now study a 3D plasma column threaded by a helical magnetic field and unstable to the current--driven kink mode. 
Both MHD and PIC numerical simulations have shown, indeed, that such configurations may naturally generate reconnection regions which can accelerate particles to non--thermal energies \citep{Striani2016, Alves2018,Bromberg2019,Davelaar2020,OrtuoMacas2022}. 
\citet[][]{Bodo13} performed an in--depth linear analysis of the instabilities and studied the development of kink instabilities that may form reconnecting regions \citep[][]{Bodo21,Bodo2021b}.

Following \cite{Bodo13}, the 3D initial configuration of the force--free magnetic field of the magnetically dominated jet is described, in cylindrical coordinates, by:

\begin{equation}
\begin{aligned}
    B_{\rm r}&= 0 \\
    B_{\varphi}&= - \frac{B_{\varphi c}}{(r /a)} \sqrt{\left[ 1 - \mathrm{exp} \left( - \frac{r^4}{a^4} \right) \right]} \\
    B_{\rm z}&= B_{\varphi c} \sqrt{ \left[ P^2_{\rm c}- \frac{\sqrt{\pi}}{a^2} \mathrm{erf} \left( \frac{r^2}{a^2} \right) \right]} \, ,
\end{aligned}
\end{equation}
where $\mathrm{erf}()$ is the error function, $a = 0.6 r_{\rm j}$ is the magnetization radius (the radius within which the magnetic field is concentrated), $r_{\rm j} = 1$ is the jet radius and $B_{\varphi c}$ is the maximum azimuthal field.
The configuration is thus characterized by the pitch angle $P_{\rm c}$ and the average hot magnetization $\sigma_{\rm h}$. 
More precisely, $P_{\rm c}$ is the value of the pitch of the magnetic field on the jet axis, defined as:

\begin{equation}
    P_{\rm c}= \left| \frac{r B_{\rm z}}{B_{\rm \varphi}} \right|_{r=0}\, ,
\end{equation}
and the hot average magnetization $\sigma_{\rm h}$ is:  

\begin{equation}
    \sigma_{\rm h} = \frac{\left\langle B^2 \right\rangle}{\rho_{\rm 0} h c^2}\, ,
\end{equation}
where $\langle B^2 \rangle = \int^a_{\rm 0} (B^2_{\rm z} + B^2_{\varphi}) r dr / \int^a_{\rm 0} r dr$ and $\sigma_{\rm h}= 10$.
The initial values of density $\rho_{\rm 0}$ and pressure $p_{\rm 0}$ are constant, with $p_{\rm 0} = 0.01 \rho_{\rm 0} c^2$ in order to have a cold jet. 
At large radii, $B_{\rm \varphi} \propto 1/r$ and $B_z$ decreases to a small constant value determined by the choice of the $P_c$ value.

Following \cite{Bodo21} we choose $P_{\rm c}/a = 1.332$ to guarantee a fast growth of the instabilities and efficient dissipation. 
For simplicity, the numerical simulations are performed in a frame in which the jet plasma is not moving ($v_{\rm z}= 0$). 

The macroparticles are initially located in the jet volume ($r<r_{\rm j}$) and, as in the 2D case, during the evolution of the simulation they can move across reconnecting regions, providing a sampling of the fluid quantities around them.

The computational box is the cube $L \times L \times L_{\rm z}$ discretized with $700 \times 700 \times 250$ grid zones, where $L = 60 r_{\rm j}$ and $L_{\rm z} = 10 r_{\rm j}$. 
The grid is uniform for $|x|$, $|y|<8$ ($x=y=0$ are on the jet axis) and geometrically stretched elsewhere in order to have a box large enough to avoid spurious effects from the lateral boundaries. 

The stretched grid is generated by incrementing the cell aspect ratio in the $x$ and $y$ directions by a factor $r$, where $r>1$ is determined by the condition that the stretched grid patch fits the domain size.
This yields the nonlinear equation 
\begin{equation}\label{eq:grid}
  r \frac{1 - r^N}{1 -r} = \frac{L/2 - 8}{\Delta x}
\end{equation}
where $\Delta x$ is the uniform mesh spacing in the inner region.
\textsc{PLUTO} solves Eq. (\ref{eq:grid}) with a standard bisection algorithm.

The boundary conditions are periodic in the $z$-direction and outflow elsewhere. 

\subsubsection{Results}

\begin{figure*}
    \centering
    \includegraphics[width=0.86\linewidth]{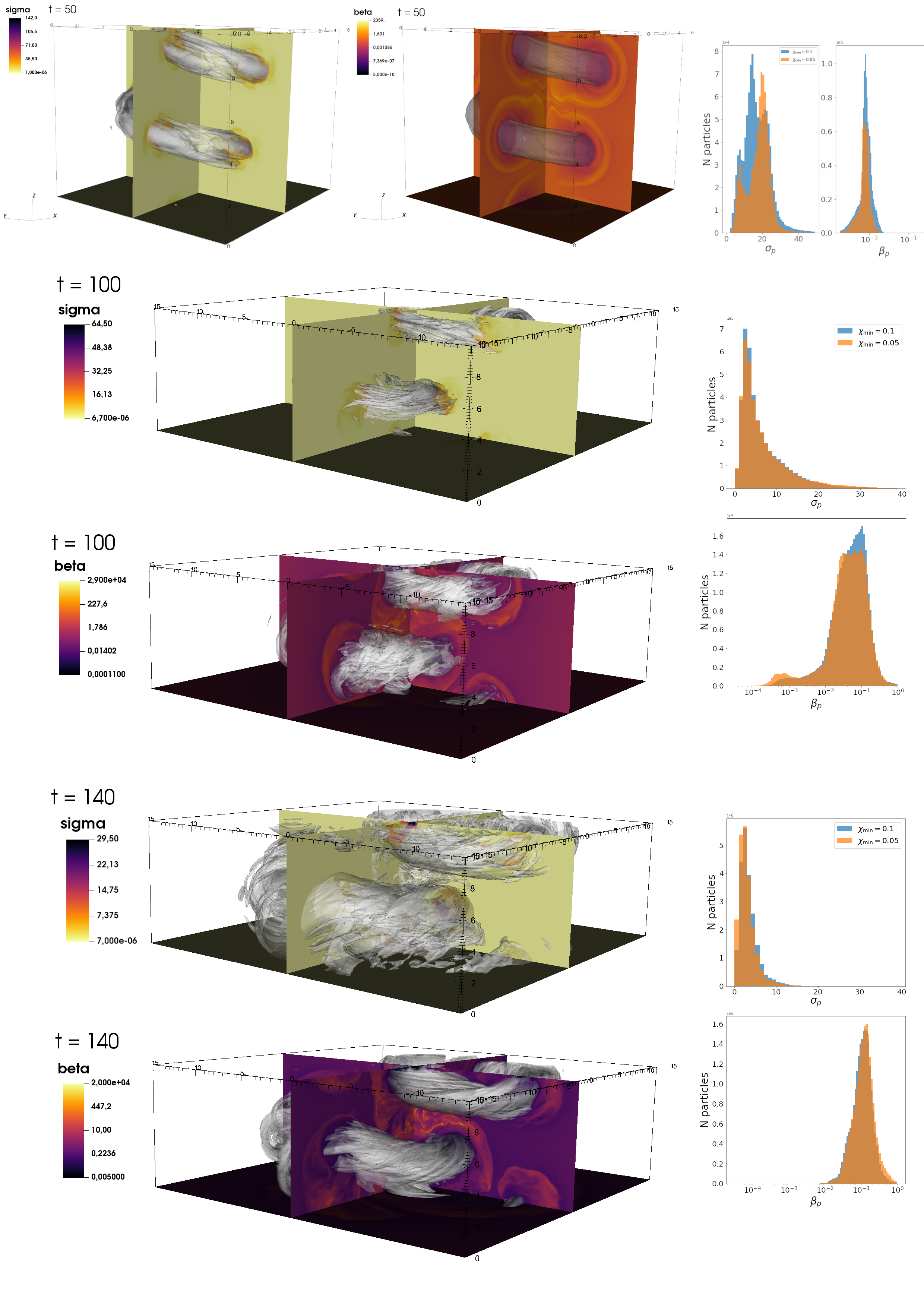}
    \caption{Results for the 3D RMHD jet simulation for an initial $\sigma_{\rm h} = 10$ and a threshold $\chi_{\rm min}=0.1$. 
    Three different times ($t=50,100,140$) are shown. 
    The domain is restricted to $x,y \in \left[-6,+6 \right]$ for $t=50$ and $x,y \in \left[-15,+15 \right]$ for $t=100,140$. 
    For each panel the jet values of $\sigma$ and $\beta$ (in logarithmic scale) are shown, with a 3D slice and an isosurface plot. 
    The right-hand panels represent (in blue) the distributions of $\sigma_{\rm p}$ and $\beta_{\rm p}$ sampled by the macroparticles that entered a reconnection region in the time interval of the respective plot time $\pm 10$. 
    The distributions coloured in orange refer to the results with a threshold $\chi_{\rm min}=0.05$.}
    \label{fig:3Djet}
\end{figure*}

\begin{figure}
    \centering
    \includegraphics[width=\linewidth]{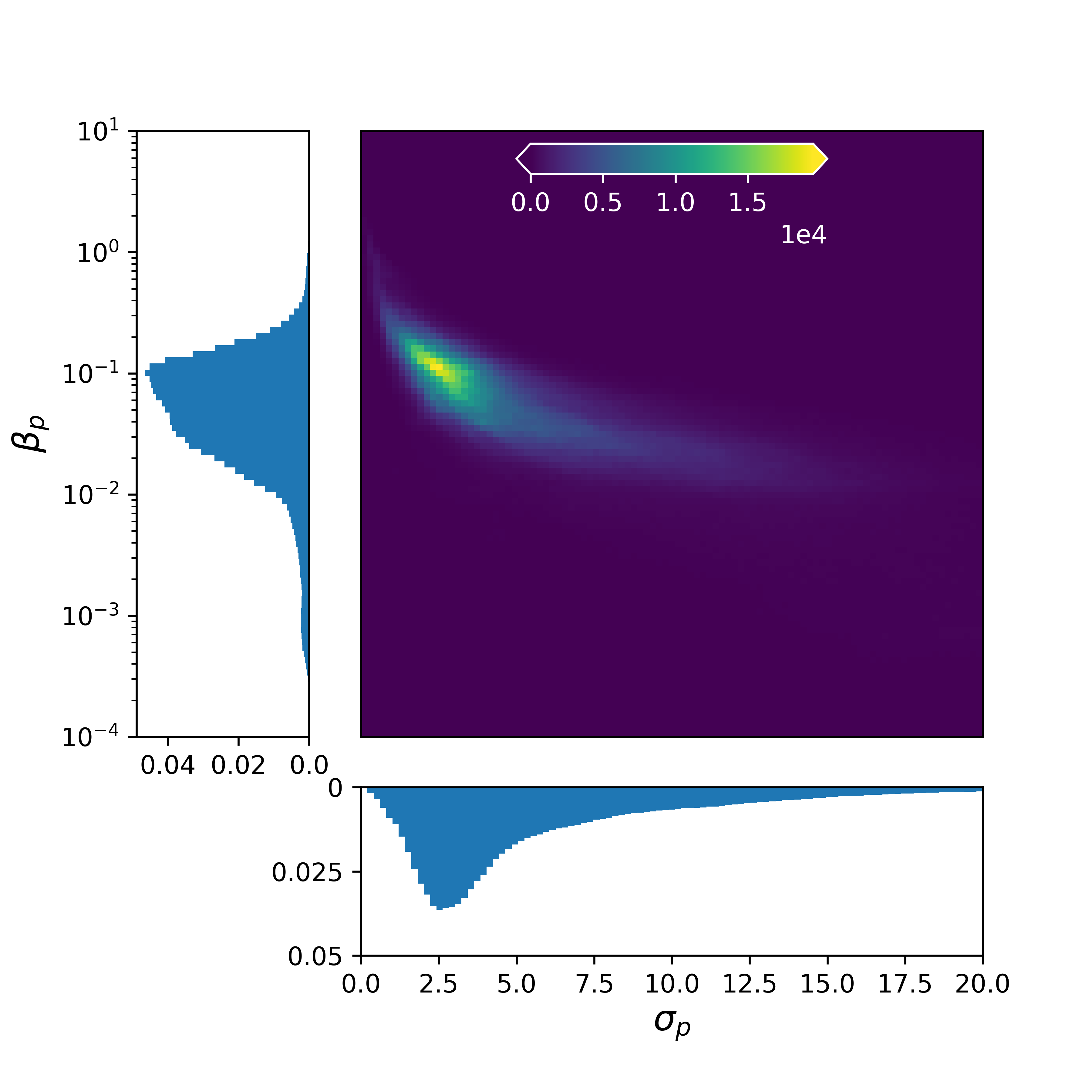}
    \caption{Same as Fig. \ref{fig:sigma_vs_beta} but for the 3D RMHD jet simulation, with initial $\sigma = 10$ at $t = 100 \pm 10$ (corresponding to $t=100$ in Fig. \ref{fig:3Djet}).
    }
    \label{fig:sigma_vs_beta3D}
\end{figure}

The results of the simulation are reported in Fig. \ref{fig:3Djet}, where $\sigma$ and $\beta$ isosurfaces are shown together with slice cuts.
We analyze 3 different simulation times $t=50, 100, 140$ in units of $r_{\rm j}/c$ representative of different phases of the evolution of the plasma column, respectively: the linear phase, the full onset of the kink instabilities and the final phase when the column gets disrupted.

A threshold $\chi_{\rm min} = 0.1$ has been set for the identification of the reconnection sites. 
Similarly to the 2D case (Sec. \ref{chap:2Djet}), the right--hand panels of the figures report the distributions (in blue color) of the sampled quantities $\sigma_{\rm p}$ and $\beta_{\rm p}$ over the whole 3D domain. 

During the linear phase ($t=50$) the reconnection sites are located at the borders of the column (its external parts), where $\sigma$ and $\beta$ achieve, respectively, large and small values.
The values of $\sigma_{\rm p}$ sampled by the macroparticles are as large as 20.
As expected, the instability leads to the formation of increasingly complex spatial structures (corresponding to the growth of different azimuthal modes, on top of the $|m| = 1$ mode) with a systematic decrease of the magnetization ($t=100,140$) and reconnection sites occupying the whole column volume. 
The distributions of $\sigma_{\rm p}$ and $\beta_{\rm p}$ at $t=50$ are still similar to the initial values and, at later times of the evolution, their values reach quasi--equipartition, with the distributions becoming more peaked around $\sigma_{\rm p}\sim 1-5$ and $\beta_{\rm p}\sim 10^{-1}$.

In order to check the effect of the threshold parameter, we also consider the case $\chi_{\rm min} = 0.05$. 
The resulting distributions of $\sigma_{\rm p}$ and $\beta_{\rm p}$ are shown in the right--hand panels of the figures using orange color.
While, during the initial phase, the distributions of $\sigma_{\rm p}$ are sensitive to the value of $\chi_{\rm min}$, at later times the distributions are basically indistinguishable, indicating that the chosen value of threshold $\chi_{\rm min} = 0.1$ could be adequate in the long--term evolution of the reconnection regions. 
We notice that a lower threshold does not increase the number of sampled particles: this is due to the fact that macroparticles tend to concentrate in the regions around reconnection sites.

We stress that in the resulting 3D structure of the reconnection regions, with very asymmetric and complex geometry of the reconnection sites, our algorithm shows its full potential for the sampling of the relevant physical parameters.

Finally, in Fig. \ref{fig:sigma_vs_beta3D}, we show again a 2D representation of the probability distributions function of $\sigma_{\rm p}$ and $\beta_{\rm p}$  at $t=100$. 
The color bar in the 2D histogram describes the number of particles with specific $\sigma_{\rm p}$ and $\beta_{\rm p}$ values. 
The vast majority of the reconnection regions are characterized by $\sigma_{\rm p} \sim 2.5$ and $\beta_{\rm p} \sim 10^{-1}$. 
PIC simulations indicate that these values are suited for efficient acceleration of a non--thermal particle distribution with typical power--law indices broadly consistent with observations. 
We notice that this differs from the results of the 2D case with  continuous injection (see above). Plausibly this is due to the greater dependence of the values sampled on the initial (and border) conditions in the slab jet with respect to the case of the evolving plasma column. 
Such (promising) findings and their consequences on the radiative emission will be explored in paper II.

\section{Summary and Conclusions}\label{chap:conclusions}

In this paper we have presented a new method to identify and characterize the physical properties of current sheets and reconnection regions in RMHD simulations, implemented in the \textsc{PLUTO} code \citep[][]{Mignone2007}. 
With respect to previous investigations, the novelty of our algorithm for the identification of reconnection sites, is the improved computational efficiency in large scale simulations, and its capability of recognizing current sheets in complex 2D and 3D geometries. 

We have tested the method in the cases of a single sheet, a slab jet and a 3D unstable plasma column, demonstrating the efficacy of the proposed method.

As our aim is to determine the effectiveness of MR events occurring in large-scale simulations in building a non--thermal distribution of accelerated particles, we have also developed an algorithm which, by means of Lagrangian macroparticles in the fluid, sample the plasma properties.
According to PIC simulations the magnetization $\sigma$ and the $\beta$ are the chief parameters which determine the efficiency, energetics and resulting particle spectra. 
Such a sampling has been performed in both the 2D and 3D simulations and the statistical properties of such parameters have been inferred.

With respect to the particle acceleration process, a limitation of the algorithm for the identification of current sheets is that, in this form, it does not directly provide a way to determine the spatial extension of the current sheets, contrary to the method proposed by \cite{Zhdankin2013}. 
This could be particularly critical if the dimensions of the reconnection sites prove to be a relevant parameter for the acceleration process \citep[see e.g.][]{Sironi16}.   
Another limitation is that the estimate of the magnetization does not take into account the possible presence of a guide field. 
A strong guide field can have a disrupting effect on the efficiency of the acceleration and the final spectrum of non--thermal particles \citep[see e.g.][]{Werner2017}. 
In principle it is possible to overcome this limitation by separating the guide field from the reconnecting one.
However a further study of the dependence of the final spectra on both the chief parameters is needed. 

Finally, a more fundamental limit could be given by the lack of energetic feedback between the fluid and the macroparticles. 
Clearly, an assessment on whether this constitutes a relevant self--consistency issue requires firstly to quantify the energetics of the non--thermal and thermal distributions of electrons and ions according to the sampled magnetization and $\beta$ parameter. 

In paper II we will focus on the acceleration of particles. 
Here we have concentrated on the implementation of the identification and sampling algorithms.
Still the simulation for the 3D case showed promising results for what concerns the sampled fluid parameters. 
Furthermore, we will also examine different conditions in terms of $\sigma$ and $\beta$ of the fluid and, adopting the prescriptions from PIC simulations, will investigate the implications on the energetics and the emitted (time dependent) radiative spectrum. In paper II several aspects will have to be examined (energetics, dynamical vs microphysical timescales, different particle species, ...). This should provide the basis for predictions and comparisons with 
astrophysical observations in a vast range of contexts (and sources) associated with relativistic jets and outflows.

\section*{Data Availability}

PLUTO is publicly available and the data will be shared
on reasonable request to the corresponding author.

\section*{Acknowledgments}

We thank the referee for the constructive feedbacks.

\bibliographystyle{mymnras}
\bibliography{biblio} 

\appendix 
\bsp	
\label{lastpage}
\end{document}